 \documentclass[12pt,preprint]{aastex}

\newcommand\tr{\textrm{\,tr\,}}

\slugcomment{.}

\shorttitle{}
\shortauthors{Kibble \& Lieu}

\begin{document}


\title{Average magnification effect of clumping of matter}


\author{T.W.B. Kibble}
\affil{Blackett Laboratory, Imperial College, London SW7
2AZ, United Kingdom\email{kibble@imperial.ac.uk}}
\and
\author{Richard Lieu}
\affil{Department of Physics, University of Alabama at
Huntsville, Huntsville, AL 35899,
U.S.A.\email{lieur@cspar.uah.edu}}




\begin{abstract}
The aim of this paper is to re-examine the question of
the average magnification in a universe with some
inhomogeneously distributed matter.  We present an
analytic proof, valid under rather general conditions,
including clumps of any shape and size, and strong lensing,
that so long as the clumps are uncorrelated the average
\emph{reciprocal} magnification (in one of several possible
senses) is precisely the same as in a homogeneous universe
with equal mean density.  From this result, we also show
that a similar statement can be made about one definition of
average \emph{direct} magnification.  We discuss, in the
context of observations of discrete and extended sources, the
physical significance of the various different measures of
magnification and the circumstances in which they are
appropriate.
\end{abstract}




\section{Introduction}

There has been considerable debate about the average
magnification effect of gravitational lensing by randomly
distributed clumps of matter.  \citet{Wei76} argued
that the average magnification produced by randomly
distributed masses is exactly the same as that in a
homogeneous universe of equal mean  (or pre-clumping) density --- the
magnification produced by the clumps is largely cancelled by
the Dyer--Roeder effect \citep{DyeR72,DyeR73}.  But his
arguments have been criticized by \citet{Ell+98}, 
who pointed out that they ignore the
effects of caustics.  These authors also introduced an
important distinction between two measures of distance,
which they called `area distance' and `angular-size
distance', though in fact both can be applied to either
lengths or areas.  \citet{HolW98} developed a
general formalism for estimating the probability
distribution of magnification, as well as shear and
rotation, and obtained numerical results for a range of
cosmological parameters using Monte Carlo simulation of
light paths.  \citet{Cla00} studied a number of
different examples, and concluded that to first order small
deviations from homogeneity would not change the average
magnification.  On the other hand, \citet{Ros01} gave an
analytic argument using a spherically symmetric model of
the universe with the aim of showing that objects in an
inhomogeneous universe appear, on average, more magnified
than those at the same redshift in a homogeneous universe
with the same mean density.  This is not in contradiction
with Claudel's result, because the effect Rose finds is of
second order.

The purpose of this paper is to re-examine this question
using a simple and explicit analytic approach.  We show that
under rather general conditions there is at least one
measure by which the average reciprocal magnification
\emph{is} exactly the same as in an FRW universe with the
same mean density.  When there is strong lensing, the
different measures of distance diverge.  It is easier to
deal initially with \emph{reciprocal} magnification, because
it goes to zero rather than infinity on the caustics. 
Later, however, we do consider average direct magnification.

Our starting point is in some respects similar to that of
\citet{HolW98}, based on using the geodesic
deviation equation to follow the paths of light signals
back in time.  Our goal is more restricted, in that we
focus only on average magnification, not rotation and
shear.  On the other hand, we are seeking analytic rather
than purely numerical estimates, so the assumptions we make
are slightly more restrictive, though still, we believe, of
wide applicability.

Specifically, we assume that in addition to a smooth,
homogeneous matter component, with density $\rho_h$, there
is another component comprising widely separated, slow
moving and randomly distributed mass clumps (say galaxies,
groups, or clusters).  For simplicity, we suppose initially
that each clump has the same mass $M$.  But it is easy to
generalize the discussion to include a distribution of
masses, even an evolving one.

Holz and Wald assumed that the universe can be described by
a `Newtonianly perturbed Friedmann--Robertson--Walker (FRW)
universe' \citep{FutS89}, \emph{i.e.}\ the metric is an FRW
metric with the time and space parts multiplied respectively
by $(1+2\phi)$ and $(1-2\phi)$, where the convention of $c=1$ is
adopted here and henceforth.
With various assumptions on
$\phi$ and the matter distribution, they showed that $\phi$
obeys a Poisson equation, with $\delta\rho=\rho-\bar\rho$
on the right hand side, where $\rho$ is the density and
$\bar\rho$ the density of the corresponding FRW universe
defined by setting $\phi=0$.  They argue that to determine
the way a light signal propagates it is sufficient to look
explicitly only at the gravitational potential of nearby
clumps.

According to our assumptions, the density perturbation
would comprise two contributions, a spatially uniform
negative background $\rho_h-\bar\rho$, and an occasional
large positive contribution from one of the clumps.  For
most of its journey, a light signal will be travelling
through a uniform background, but when it does pass near
a clump the effects will be much larger.  Under these
conditions it is reasonable to assume that we can deal
with the effects of the clumps individually.  We assume
that the clumps are small and slow-moving enough that the
gravitational effect of each one may be treated in a
Newtonian approximation, with a time-independent Newtonian
potential $\Phi$.  Moreover, we use the `plane lens
approximation', that is, we compute the angle of deviation
due to the clump by integrating the gradient of the potential
along the undeviated light path, and assume that the
deviation effectively occurs at the central plane.  As
pointed out by \citet{MetS97}, this induces a small error
because the true light path passes closer to the center.
However the discrepancy in the minimum distance from the
center is very small, of order the Schwarzschild radius of
the clump.  Hence the error is tiny and consistently
negligible in the Newtonian approximation.  Finally, we also
assume that far from the clumps there is no appreciable
source of shear, so that the Weyl tensor vanishes.  Of
course, no such assumption is made about the field near each
clump.

One criticism that might be made is of our assumption that
the clumps are well separated and randomly distributed. 
This does not mean however that only one clump can
significantly affect a light signal at any time (though
that may often be true), but rather that the effects of
different clumps are purely additive.  This seems to us
generally a good approximation.  The most serious objection
would probably be to the assumption that the clumps are
\emph{uncorrelated}.  Such correlations may invalidate the
assumption that there is no source of shear far from the
clumps.  Even in such cases, the effect on average
magnification should be small, since according to the
Raychaudhuri equation the effect of shear on expansion is
of second order.  These correlations might also be thought
to call in question the validity of the plane-lens
approximation, but this would be true only if the clumps are
correlated in such a way that the deviated light paths
sample a significantly different environment.  Given the
extremely small error in the deviation angle $\psi$
(typically of order $\psi^2$), this seems very unlikely.

It is important to note that the `average magnification'
for a given red-shift can mean several different things. 
In the strong-lensing case, when caustics are present,
imaged areas fold back on themselves.  In one sense, the
magnification is \emph{negative} in the region beyond the
caustic, because images are reversed.  In the distinction
made by \citet{Ell+98}, in computing the
\emph{angular-size distance}, these regions are indeed
counted negatively, whereas the \emph{area distance} is
concerned with the total area, including all the folds; in
that case, every contribution is taken positively.

There is another important distinction to be made.  We
may choose at random one of the sources at red-shift $z$,
or we may choose a random direction in the sky and look for
sources there.  These are not the same; the choices are
differently weighted.  If one part of the sky is more
magnified, or at closer angular-size distance, the
corresponding area of the constant-$z$ surface will be
smaller, so fewer sources are likely to be found there.  In
other words, choosing a source at random will give on
average a smaller magnification or larger angular-size
distance.

Which of these definitions is appropriate depends on what we
choose to look at, and what questions we want to ask.  We
shall return to the question of which definition to use in
various circumstances in a later section.

Let us concentrate for the moment on the random-direction
averaging.  The question we wish to address is this: How is
the average magnification affected by whether the matter is
clumped rather than smoothly distributed?  We do this by
examining the geodesic deviation equation in the presence
of clumps. 

One other preliminary point should be made.  What we are
interested in observationally is the average magnification
of sources at a given red-shift $z$.  But what we actually
calculate is the average of sources at the same affine
distance $\lambda$ (along the backward null geodesics from
the present), which is not exactly the same thing.  We
argue however that the difference is undetectably small. 
The effect of passing near a clump of mass $M$ affects the
relationship between $z$ and $\lambda$ in much the same way
as the conventional gravitational time delay.  Thus the
difference in $z$ for fixed $\lambda$ is of order $H_0GM$
times a logarithmic factor, which is negligible under any
reasonable conditions.

\section{Null geodesics}

The Robertson--Walker line element for an open universe,
with $k=-|k|$ and $c=1$, is
 \begin{equation}
ds^2=dt^2-a^2(t)\left\{\frac{dr^2}{1+|k|r^2}
+r^2(d\vartheta^2+\sin^2\vartheta\,d\varphi^2)\right\},
 \end{equation}
or equivalently, with $\tau=\int dt/a(t)$ and
$r=|k|^{-1/2}\sinh(|k|^{1/2}\chi)$,
 \begin{equation}
ds^2=a^2(\tau)\left\{d\tau^2-d\chi^2-\frac{1}{|k|}
\sinh^2(|k|^{1/2}\chi)(d\vartheta^2
+\sin^2\vartheta\,d\varphi^2)\right\}.
 \end{equation}
Of course in the flat-space limit, $|k|\to0$, $r$ and
$\chi$ become identical.

The Friedmann equation is
 \begin{equation}
H^2=\frac{1}{a^2}\left(\frac{da}{dt}\right)^2=
\frac{8\pi G}{3}\rho_m+\frac{|k|}{a^2}+\frac{\Lambda}{3},
 \end{equation}
where $\rho_m$ is the density of matter (assumed
pressureless).  Consequently, the relation between the
Hubble parameter $H$ and the red-shift, $z=a_0/a(t)-1$, is
$H=H_0E(z)$, where
 \begin{equation}
E^2(z)=\Omega_m(1+z)^3+(1-\Omega_m-\Omega_\Lambda)(1+z)^2
+\Omega_\Lambda,
 \end{equation}
in which as usual $\Omega_m=8\pi G\rho_{m0}/3H_0^2$ and
$\Omega_\Lambda=\Lambda/3H_0^2$.

We shall consider backward null geodesics from the origin
at the present time $t_0$, with affine parameter $\lambda$
normalized so that $\dot t(0)=-1$, where the dot denotes a
derivative with respect to $\lambda$.  
Then
 \begin{equation}
-\dot\tau(\lambda)=\dot\chi(\lambda)
=\frac{[1+z(\lambda)]^2}{a_0}.
\label{chidot}
 \end{equation}

We now assume that in addition to a uniform distribution of
matter there are random clumps present.  Specifically, the
matter density parameter $\Omega_m$ may be written
$\Omega_m=\Omega_h+\Omega_g$, where $\Omega_h$ represents a
homogeneous distribution and $\Omega_g$ a random
distribution of widely separated clumps, each of mass $M$. 
(It is easy to generalize the discussion to a distribution
of masses, or even to allow for a distribution changing
with cosmic time.)

Consider a fiducial backward null geodesic, and a second
neighboring null geodesic from the same point.   We choose
a vierbein $e_{(\mu)}$ at the origin, with $e_{(0)}$ in the
$t$ direction, and $e_{(3)}-e_{(0)}$ tangent to the
fiducial geodesic.  Then we parallel-propagate the vierbein
along this geodesic, and introduce transverse coordinates
$\vec l=(l^1,l^2)$, such that the transverse separation
between the geodesics at affine distance $\lambda$ is 
 \begin{equation}
\delta x^\mu=l^{\alpha}e_{(\alpha)}^\mu(\lambda).
 \end{equation}
(Summation over $\alpha=1,2$ is implied.)  The rate of
change of this separation is governed by the geodesic
deviation equation,
 \begin{equation}
\ddot l^{\alpha}=
R^\alpha{}_{\mu\beta\nu}u^\mu u^\nu l^\beta,
\label{geodev}
 \end{equation}
where $u^\mu=\dot x^\mu$ is the null tangent vector to the
fiducial geodesic.

It is convenient to define the $2\times2$ matrix ${\cal
A}(\lambda)=\Bigl({\cal A}^\alpha{}_\beta(\lambda)\Bigr)$ by
 \begin{equation}
\vec l(\lambda)={\cal A}(\lambda)\cdot\dot{\vec l}(0),
\label{Adef}
 \end{equation}
where $\dot{\vec l}(0)$ defines the initial (small) angular
deviation from the fiducial geodesic.  Then ${\cal A}$ also
satisfies the geodesic deviation equation (\ref{geodev}). 
It is also useful to introduce the quantity
 \begin{equation}
A(\lambda)=\det{\cal A}(\lambda),
\label{detA}
 \end{equation}
which is proportional to the reciprocal magnification of a
small source. In the weak-lensing case, $A$ is always
positive.  Then the angular-size (or area) distance $D$ is
given by $D=\sqrt{A}$.  However, caustics are lines of
infinite magnification, where $A=0$.  In the regions
beyond, where $A<0$, images are reversed.  More generally,
 \begin{equation} 
D(\lambda)=\sqrt{|A|}.
\label{angdiam}
 \end{equation}
For a single image of a point source, there is no need to
distinguish angular-size and area distances.  The
distinction for large sources is discussed in Section 5.

So long as we are well away from the clumps of matter, it
is reasonable to assume that the Weyl tensor is very small,
\emph{i.e.}, there is no source of shear.  We make no such
assumption about the effect of individual clumps.  (Each
clump of course introduces a source of shear in its
vicinity.)  Then the coefficient on the right hand
side of the geodesic deviation equation involves only the
Ricci tensor.  Using Einstein's equations, this may be
expressed in terms of the density $\rho_h$ of the smooth
component, or equivalently $\Omega_h$.  Thus, along light
paths that come near to none of the clumps, we find for
$\cal A$ the equation
 \begin{equation}
\ddot{\cal A}=-\frac{3}{2}H_0^2\Omega_h(1+z)^5{\cal A}.
\label{geodevA}
 \end{equation}
Here a factor $(1+z)^3$ comes from the $z$-dependence of
$\rho_h$, while $(1+z)^2$ derives from the $u^\mu u^\nu$
factor.  Clearly, away from the clumps, $\cal A$ is the
solution of this equation (\ref{geodevA}) with initial
conditions
 \begin{equation}
{\cal A}(0)={\bf 0},\qquad\dot{\cal A}(0)={\bf 1}.
 \end{equation}

Thus for these paths, we find ${\cal 
A}(\lambda)=D_h(\lambda){\bf 1}$, where $D_h(\lambda)$
is the Dyer--Roeder distance, the solution of the equation
 \begin{equation}
\ddot D_h=-\frac{3}{2}H_0^2\Omega_h(1+z)^5D_h,
\label{Dh}
 \end{equation}
with initial conditions
 \begin{equation}
D_h(0)=0,\qquad\dot D_h(0)=1.
\label{Dhic}
 \end{equation}

It will also be useful to define the corresponding
angular-diameter distance $\bar D(\lambda)$ for a
homogeneous FRW universe with density $\bar\rho$, which is
in fact given by 
 \begin{equation}
\bar D(\lambda)=a(\lambda)r(\lambda).
\label{barD}
 \end{equation}

If the relative magnification $\mu$ of a point source,
compared to that of a source at the same red-shift in the FRW
universe is defined to be \emph{negative} when the image is
reversed, then 
 \begin{equation}
\mu=\frac{\bar A(\lambda)}{A(\lambda)}.
\label{mu}
 \end{equation}
The magnification in the more usual sense, which is always
positive, is then
 \begin{equation}
|\mu|=\frac{\bar A(\lambda)}{|A(\lambda)|}
=\frac{\bar D^2(\lambda)}{D^2(\lambda)}.
 \end{equation}

\section{Effect of a clump}

We now consider the effect of a clump near our null
geodesics at affine distance $\lambda$.  We suppose that
the center of the clump is at transverse position $-\vec x$
from the fiducial geodesic.  The effect of the clump will be
to bend the geodesic through an angle $\vec\psi$.  With the
center of the clump as origin, let us introduce Euclidean
coordinates $(\vec x,\zeta)$.  Then $\vec\psi$ is given in
terms of the Newtonian potential $\Phi(\vec x,\zeta)$ of
the clump by
 \begin{equation}
\vec\psi(\vec x)=
-2\int_{-\infty}^{\infty}d\zeta\,
\vec\nabla\Phi(\vec x,\zeta),
\label{psidef}
 \end{equation}
where $\vec\nabla$ is the two-dimensional gradient
operator.  Note that if $\rho(\vec x,\zeta)$ is the density
distribution in the clump, then, by virtue of the
Poisson equation,
 \begin{equation}
\vec\nabla\cdot\vec\psi(\vec x)=4\pi G\sigma(\vec x),
\label{Poiss}
 \end{equation}
with
 \begin{equation}
\sigma(\vec x)=\int_{-\infty}^{\infty}d\zeta\,
\rho(\vec x,\zeta).
 \end{equation}
 
The angular deviation of a neighboring geodesic at
transverse displacement $\vec l$, relative to that of the
fiducial geodesic, will be 
 \begin{equation}
\vec\psi(\vec x+\vec l)-\vec\psi(\vec x)=
{\cal K}(\vec x)\cdot\vec l,
 \end{equation}
where the $2\times2$ matrix $\cal K$ is given by
 \begin{equation}
{\cal K}(\vec x)=\vec\nabla\vec\psi(\vec x).
\label{Kdef}
 \end{equation}
Note that by the form of (\ref{psidef}) $\cal K$ is
symmetric.

Now from (\ref{chidot}) and (\ref{Adef}), the angular
deviation of neighboring geodesics at position $\lambda$ is
 \begin{equation}
\frac{\dot{\vec l}}{a\dot\chi}=\frac{\dot{\vec l}}{1+z}=
\frac{\dot{\cal A}\cdot\dot{\vec l}(0)}{1+z}.
 \end{equation}
Hence $\dot{\cal A}$ changes sharply as the geodesic passes
the clump by the amount
 \begin{equation}
\delta\dot{\cal A}=
(1+z){\cal K}(\vec x)\cdot{\cal A}(\lambda).
\label{deltacalAdot}
 \end{equation}

It is also possible to find a similar expression for
$\delta\dot A$.  Since 
 \begin{equation}
\dot A=A\,\tr({\cal A}^{-1}\dot{\cal A}),
 \end{equation}
it follows that
 \begin{equation}
\delta\dot A=A\,\tr({\cal A}^{-1}\delta\dot{\cal A})
=(1+z)A\,\tr{\cal K}(\vec x).
\label{deltaAdot}
 \end{equation}

Our aim now is to compute the expected value of the change
in $\dot{\cal A}$ in a small interval $d\lambda$ due to the
effect of a random distribution of clumps.

Since the clumps are of mass $M$ and constitute a fraction
$\Omega_g$ of the critical density, their number density
(assuming no evolution) is clearly
 \begin{equation}
n=n_0(1+z)^3,\qquad n_0=\frac{3H_0^2\Omega_g}{8\pi GM}.
\label{no}
 \end{equation}
The probability of finding a clump with center at the
position $(\lambda,-\vec x)$ within small ranges
$d\lambda,d^2\vec x$ is
 \begin{equation}
n_0(1+z)^4\,d\lambda\,d^2\vec x,
\label{prob}
 \end{equation}
where the extra factor of $(1+z)$ occurs because
$ad\chi=(1+z)d\lambda$. 

To find the equation we are looking for, let us ask how
$\langle\dot{\cal A}\rangle$ changes under a small change
$d\lambda$ in $\lambda$.  Of course, due to the smooth
background, according to (\ref{geodevA}), there is a change
 \begin{equation}
[d\langle\dot{\cal A}(\lambda)\rangle]_h=
-\frac{3}{2}H_0^2\Omega_h(1+z)^5\langle{\cal
A}(\lambda)\rangle d\lambda.
\label{Adot_h}
 \end{equation}

Now what is the average effect of the clumps?  The
probability that a clump is present is given by
(\ref{prob}).  Since we have assumed the clumps are
uncorrelated, this probability is of course independent of
the previous history, so by (\ref{deltacalAdot}) the change
due to clumps is 
 \begin{equation}
[d\langle\dot{\cal A}(\lambda)\rangle]_g=
n_0(1+z)^5 d\lambda\int d^2\vec x\,{\cal K}(\vec
x)\cdot\langle{\cal A}(\lambda)\rangle.
\label{Adot_g}
 \end{equation}

The next step is to compute the $\vec x$ integral.  Using
(\ref{Kdef}), we have
 \begin{equation}
\int d^2\vec x\,{\cal K}(\vec x)=
\int d^2\vec x\,\vec\nabla\vec\psi(\vec x)=
\oint dl\,\vec n\vec\psi(\vec x),
 \end{equation}
where $\vec n=\vec x/|\vec x|$, and the integral is around a
large circle.  Now for large
$\vec x$ we have asymptotically
 \begin{equation}
\vec\psi(x)\sim-4GM\frac{\vec x}{\vec x^2},
 \end{equation}
whence
 \begin{equation}
\int d^2\vec x\,{\cal K}(\vec x)=-4\pi GM{\bf 1}.
\label{intK}
 \end{equation}

For later use, we need one other result, concerning the
integral of the determinant of ${\cal K}$.  If
$\hat\zeta$ is a unit vector in the direction of the
fiducial null ray, then clearly
 \begin{equation}
\hat\zeta\det{\cal K}=
\vec\nabla\psi_1\times\vec\nabla\psi_2=
\vec\nabla\times(\psi_1\vec\nabla\psi_2).
 \end{equation}
Hence
 \begin{equation}
\int d^2\vec x\,\det{\cal K}=
\oint d\vec l\cdot(\psi_1\vec\nabla\psi_2),
 \end{equation}
where the integral is again around a large circle.  Clearly
this vanishes in the limit, whence
 \begin{equation}
\int d^2\vec x\,\det{\cal K}=0.
\label{intdetK}
 \end{equation}

Now it follows from (\ref{Adot_g}), (\ref{no}) and
(\ref{intK}) that
 \begin{equation}
[d\langle\dot{\cal A}(\lambda)\rangle]_g=
-\frac{3}{2}H_0^2\Omega_g(1+z)^5\langle{\cal
A}(\lambda)\rangle d\lambda.
 \end{equation}

Remarkably enough, this is of precisely the same form as
(\ref{Adot_h}), so, combining the two, we find
 \begin{equation}
\langle\ddot{\cal A}(\lambda)\rangle=
-\frac{3}{2}H_0^2\Omega_m(1+z)^5
\langle{\cal A}(\lambda)\rangle.
\label{Addot}
 \end{equation}
In other words, $\langle{\cal A}(\lambda)\rangle$ is
exactly the same as it would be in a homogeneous universe
of equal mean density, namely
 \begin{equation}
\langle{\cal A}(\lambda)\rangle=\bar D(\lambda){\bf 1},
\label{AbarD}
 \end{equation}
where $\bar D$ is given by (\ref{barD}).

It is not possible to use (\ref{deltaAdot}) directly to find
a similar equation for the mean value of the quantity $A$
defined by (\ref{detA}), because even in the homogeneous
background $A$ does not satisfy a simple homogeneous
differential equation like (\ref{geodevA}).  Nevertheless,
we can also show that
 \begin{equation}
\langle A(\lambda)\rangle=[\bar D(\lambda)]^2.
\label{detAbarD}
 \end{equation}
The general proof of this result is rather long and is
therefore relegated to an appendix.  However, in the next
section we explain why it is true in the special case where
a typical light path encounters no more than one clump.

This result, (\ref{detAbarD}), shows that in a clumpy
universe the average \emph{reciprocal} magnification
$\langle 1/\mu\rangle$, where $\mu$ is defined by
(\ref{mu}), is exactly the same as in a homogeneous
universe of equal mean density.

It is remarkable that, despite the quadratic relationship
between $\cal A$ and $A$, the averages of both,
$\langle{\cal A}\rangle$ and $\langle A\rangle$, can be
expressed in terms of this one function $\bar D$.  It is
very important to note, however, that $\bar D$ is
\emph{not} the average $\langle D\rangle$ of the
angular-size or area distance of point sources --- and
nor is $\bar D^2$ equal to $\langle D^2\rangle$ --- because
according to (\ref{angdiam}), $\langle D^2\rangle=\langle
|A|\rangle$.  As we shall see, for a large source, $\bar
D$ is approximately the `angular-size distance' of 
\citet{Ell+98}, not the `area distance'.

\section{The single-clump case}

To clarify the reason for the important result
(\ref{detAbarD}) we shall in this section specialize to the
single-clump case, where the effect of multiple encounters
is assumed to be negligible.  This will make it easier to
estimate the size of the differences.

Let us begin by finding a more explicit expression for
$\langle{\cal A}(\lambda_s)\rangle$ for sources at affine
distance $\lambda_s$ in the case where the probability of
encountering more than one clump is negligible.  It will be
useful to extend the definition of the angular-diameter
distance in the `empty' regions, given by (\ref{Dh}) and
(\ref{Dhic}).  We define
$D_h(\lambda_1,\lambda)$ to be the angular-diameter
distance between $\lambda_1$ and $\lambda$ in the smooth
background between the clumps, \emph{i.e.}, the solution of
(\ref{Dh}) with initial conditions
 \begin{equation}
D_h(\lambda_1,\lambda_1)=0,\qquad
\frac{\partial}{\partial\lambda}D_h(\lambda_1,\lambda)
\bigg|_{\lambda=\lambda_1}=1+z_1.
\label{Dh2ic}
 \end{equation}
(The factor of $(1+z_1)$ appears because $a\dot\chi=1+z$.)

If there are no clumps near the light path, then of course
${\cal A}(\lambda_s)=D_h^2(\lambda_s){\bf 1}$.  If there is
a single clump at $\lambda$ and transverse position $-\vec
x$, then its deviating effect is added, so one finds
 \begin{equation}
{\cal A}(\lambda_s)=D_h(\lambda_s){\bf 1}+
D_h(\lambda,\lambda_s){\cal K}(\vec x)D_h(\lambda),
 \end{equation}
or equivalently
 \begin{equation}
{\cal A}(\lambda_s)=D_h(\lambda_s)[{\bf 1}+
L(\lambda,\lambda_s){\cal K}(\vec x)],
\label{A1cl}
 \end{equation}
where
 \begin{equation}
L(\lambda,\lambda_s)=\frac{D_h(\lambda)
D_h(\lambda,\lambda_s)}{D_h(\lambda_s)}.
 \end{equation}
It will be useful to define the quantity
 \begin{equation}
J=\det[{\bf 1}+L(\lambda,\lambda_s){\cal K}(\vec x)],
\label{J}
 \end{equation}
so that
 \begin{equation}
A(\lambda_s)=D_h^2(\lambda_s)J,
\label{AJ}
 \end{equation}
Thus we see that $J$ may be interpreted as the contribution
of this clump to the reciprocal magnification (relative to
the `empty' regions).  Note that $J$ may be negative; it
vanishes on the caustics.  Note also that
 \begin{equation}
J=1+L(\lambda,\lambda_s)\tr{\cal K}+
L^2(\lambda,\lambda_s)\det{\cal K}.
\label{Jexp}
 \end{equation}

To find the average $\langle A(\lambda_s)\rangle$, we have
to mutiply (\ref{AJ}) by the probability (\ref{prob}) of
finding a clump at position $(\lambda,-\vec x)$ and
integrate over $\lambda$ and $\vec x$:
 \begin{equation}
\langle A(\lambda_s)\rangle=D_h^2(\lambda_s)\left[1+\int
d\lambda\,n_0(1+z)^4\int d^2\vec x(L\tr{\cal
K}+L^2\det{\cal K})\right].
\label{detA1cl}
 \end{equation}
Using (\ref{no}), (\ref{intK}) and (\ref{intdetK}), we then
find
 \begin{equation}
\langle A(\lambda_s)\rangle=D_h^2(\lambda_s)\left[1-
3H_0^2\Omega_g\int
d\lambda\,(1+z)^4L(\lambda,\lambda_s)\right],
\label{detA1cl2}
 \end{equation}
which to this order is just (\ref{detAbarD}).  Note the
crucial importance in deriving this result of the fact that
the integral of the determinant of ${\cal K}$ vanishes.

\section{The different averages}

To understand the important distinction between
\emph{angular-size} and \emph{area} distances or
magnifications, it is helpful to consider the shape of the
surface of constant $z_s$ (or constant $\lambda_s$, which as
we argued is nearly the same thing) on the backward light
cone from the present.  When only weak lensing occurs, it is
obviously very close to being a sphere.  When there are
caustics, the light cone folds back and intersects itself. 
Then the world line of a comoving source will meet the cone
several times, leading to multiple images.  But it is
important to realise that the different sheets of the
surface are very close to each other; the time delay
between the different images is always very small on a
comsological scale.  So the constant-$z_s$ surface still lies
very close to a sphere; it can essentially be described as a
single sphere with some sections covered several times.  We
shall call this the \emph{source sphere}.

Now let us partition the celestial sphere around the
observer into small pixels, labelled by $n$, with solid
angles $\delta\Omega_n$ (not necessarily all equal), small
enough that the magnification may be taken to be constant
within each.  Of course $\sum_n\delta\Omega_n=4\pi$.
For each $n$ the light arriving in the $n$th pixel
originated from some small area $\delta S_n$ of the source
sphere, with
 \begin{equation}
\delta S_n=A_n(\lambda_s)\delta\Omega_n.
 \end{equation}
Note that some of these areas may be negative.

Now, the result (\ref{detAbarD}) established above shows that
the total area of the source sphere is
 \begin{equation}
\sum_n\delta S_n=\sum_nA_n(\lambda_s)\delta\Omega_n
=4\pi\langle A(\lambda_s)\rangle=4\pi\bar D^2(\lambda_s).
 \end{equation}
With the demonstration that the area of the source sphere is the
same as in the FRW universe of equal mean density, we provided a
rigorous proof of Weinberg's argument, which was based on the principle
of energy conservation \citep{Wei76}. 

We can therefore define the solid angle
$\overline{\delta\Omega}_n$ that would be subtended by
$\delta S_n$ in the homogeneous FRW universe by
 \begin{equation}
\overline{\delta\Omega}_n=\frac{\delta S_n}{\bar
D^2(\lambda_s)}.
 \end{equation}
Then the magnification $\mu_n$ relative to the FRW universe
is
 \begin{equation}
\mu_n=\frac{\delta\Omega_n}{\overline{\delta\Omega}_n}
=\frac{\bar D^2(\lambda_s)}{A_n(\lambda_s)}.
 \end{equation}
while the reciprocal magnification is
 \begin{equation}
\frac{1}{\mu_n}
=\frac{\overline{\delta\Omega}_n}{\delta\Omega_n}
=\frac{A_n(\lambda_s)}{\bar D^2(\lambda_s)}.
 \end{equation}

The \emph{random-direction} average of any quantity $f$
defined on the celestial sphere is given by
 \begin{equation}
\langle f\rangle_{\mathrm{rd}}
=\sum_nf_n(\lambda_s)\delta\Omega_n.
 \end{equation}
In particular, of course,
 \begin{equation}
\left\langle\frac{1}{\mu}\right\rangle_{\mathrm{rd}}
=\frac{1}{4\pi}\sum_n\frac{\delta\Omega_n}{\mu_n}
=\frac{1}{4\pi}\sum_n\overline{\delta\Omega}_n=1.
\label{avrecipmag}
 \end{equation}

Note that some of the terms in (\ref{avrecipmag}) are
negative.  If we consider instead the average of the modulus,
we get
 \begin{equation}
\left\langle\frac{1}{|\mu|}\right\rangle_{\mathrm{rd}}
=\frac{1}{4\pi}\sum_n|\overline{\delta\Omega}_n|>1.
 \end{equation}
In fact, multiplied by $\bar D^2(\lambda_s)$ this gives the
\emph{full} area of the source sphere, counting the regions
covered more than once with the appropriate multiplicity. 
We may define the \emph{areal}, as opposed to
\emph{angular-size}, distance $\bar D_{\mathrm{area}}$ by
 \begin{equation}
\bar D^2_{\mathrm{area}}=\bar D^2(\lambda_s)
\left\langle\frac{1}{|\mu|}\right\rangle_{\mathrm{rd}}.
\label{Darea}
 \end{equation}
Obviously, $\bar D_{\mathrm{area}}(\lambda_s)>\bar
D(\lambda_s)$.

We can also discuss \emph{random-source} averages.  To
do that it is conveneient to choose the pixels in a special
way.  Let us first partition the \emph{source sphere} into
elements labelled by $p$, of area $\bar
D^2(\lambda_s)\overline{\delta\Omega}_p$.  Here all
$\overline{\delta\Omega}_p$ are positive, and
$\sum_p\overline{\delta\Omega}_p=4\pi$.  In general, each
element may have several different images, though always an
odd number, say $2r_p+1$.  Let us label these image pixels
$\delta\Omega_{pq}$, with $q=1,2,\dots,2r_p+1$.  If we
choose $q$ so that positive and negative images alternate,
then
$\overline{\delta\Omega}_{p1}=\overline{\delta\Omega}_p$,
$\overline{\delta\Omega}_{p2}=-\overline{\delta\Omega}_p$,
\dots, and in general
 \begin{equation}
\overline{\delta\Omega}_{pq}=
(-)^{q-1}\overline{\delta\Omega}_p.
 \end{equation}
It then follows that
 \begin{equation}
\sum_q\overline{\delta\Omega}_{pq}=4\pi,
 \end{equation}
but
 \begin{equation}
\sum_q|\overline{\delta\Omega}_{pq}|=
4\pi\frac{\bar D^2_{\mathrm{area}}(\lambda_s)}{\bar
D^2(\lambda_s)}.
 \end{equation}

The random-source average of some quantity $f$ defined for
each element of the \emph{source} sphere is
 \begin{equation}
\langle f\rangle_{\mathrm{rs}} = \frac{1}{4\pi} \sum_p f_p
\overline{\delta\Omega}_p.
 \end{equation}
Consider for example $N$ sources of given absolute magnitude
randomly distributed on the source sphere, with
$n_p\overline{\delta\Omega}_p$ in element $p$.  Then clearly
 \begin{equation}
\langle n\rangle_{\mathrm{rs}}=\frac{N}{4\pi}.
 \end{equation}
What can one say about their average magnification? 
Actually, there are two possible meanings of the term.  The first is
appropriate if the various images of each source are
\emph{unresolved}.  Then the total magnification of a source
in $\overline{\delta\Omega}_p$ is the sum of the
magnifications of the separate images, each counted
positively,
 \begin{equation}
\mu_p^{\mathrm{tot}}=\sum_q|\mu_{pq}|.
 \end{equation}
It follows that
 \begin{equation}
\langle \mu^{\mathrm{tot}}\rangle_{\mathrm{rs}}=
\frac{1}{4\pi}\sum_{p,q}|\mu_{pq}|\overline{\delta\Omega}_p=
\frac{1}{4\pi}\sum_{p,q}|\mu_{pq}|
|\overline{\delta\Omega}_{pq}|.
\label{mutotav1}
 \end{equation}
However, the signs of $\mu_{pq}$
and $\overline{\delta\Omega}_{pq}$ are the same, so we may
remove the modulus signs, to obtain
 \begin{equation}
\langle \mu^{\mathrm{tot}}\rangle_{\mathrm{rs}}=
\frac{1}{4\pi}\sum_{p,q}\delta\Omega_{pq}=1.
\label{mutotav}
 \end{equation}
Thus from our earlier result, that the
\emph{random-direction} average of \emph{reciprocal}
magnification in unity, we have an important corolla ry, that
the \emph{random-source} average of \emph{total, direct}
magnification is also unity.  Note, however, the important
distinction, that every term in (\ref{mutotav1}) is
positive, whereas many terms in (\ref{avrecipmag}) are
negative.

The situation is different if the various images are
\emph{resolved} (as may usually happen when the lenses are
clusters).  Then the quantity of interest would be the
average magnification of the individual images.  Now clearly
the total amount of radiant energy arriving at the observer
is the same whether the images are resolved or not.  The difference
is that the total number of images is larger than
the number of sources.  To define the random-source average
for \emph{resolved} images, we should divide not by $4\pi$
but by the total solid angle the source sphere
with all its folds subtend, namely $4\pi\bar D^2_{\mathrm{area}}/\bar
D^2=4\pi\langle|\mu|^{-1}\rangle_\mathrm{rd}$.  Hence
 \begin{equation}
\langle \mu^{\mathrm{resolved}}\rangle_{\mathrm{rs}}=
\frac{\bar D^2(\lambda_s)}
{\bar D^2_{\mathrm{area}}(\lambda_s)}
=\frac{1}{\langle|\mu|^{-1}\rangle_\mathrm{rd}}.
\label{muresav}
 \end{equation}
Here again we have a relation between a random-source and a
random-direction average.

We conclude this section by estimating the magnitude of
the difference between areal and angular-size distances for
the single-clump approximation used in the preceding
section.  The reciprocal magnification $1/\mu$ when the
light passes a single clump at $(\lambda,-\vec x)$ is
 \begin{equation}
\frac{1}{\mu}=
\frac{D_h^2(\lambda_s)}{\bar D^2(\lambda_s)}J,
 \end{equation}
where $J$ is given by (\ref{J}).  In going from the
angular-size to the areal distance, according to
(\ref{Darea}), we have to replace $\mu$ by $|\mu|$, which in
this case amounts to replacing $J$ by $|J|$.  Since the
integrand in (\ref{detA1cl}) is just $J-1$, this replacement
yields
 \begin{equation}
\bar D^2_{\mathrm{area}}=\langle|A(\lambda_s)|\rangle=
D_h^2(\lambda_s)\left[1+
\int d\lambda\,n_0(1+z)^4\int d^2\vec x\,(|J|-1)\right].
 \end{equation}

To be specific, we adopt essentially the modified isothermal
sphere profile suggested by \citet{Bra+96}.  In this
distribution, the density at radial distance $r$ is given by
 \begin{equation}
\rho(r)=\frac{\sigma^2}{2\pi G}\frac{s^2}{r^2(r^2+s^2)},
\label{denprof}
 \end{equation}
where $\sigma$ is the line-of-sight velocity dispersion, and
$s$ is a truncation scale, beyond which $\rho$ falls
rapidly, like $1/r^4$.  The total mass $M$ of this
distribution is finite, with
 \begin{equation}
M=\frac{\pi\sigma^2 s}{G}.
 \end{equation}
For our purposes it is more convenient to adopt a slightly
modified version of the profile, designed to avoid the
infinite density at $r=0$.  We take
 \begin{equation}
\rho(r)=\frac{\sigma^2}{2\pi
G}\frac{s^2}{(r^2+a^2)(r^2+s^2)},
\label{denprof2}
 \end{equation}
where the core radius $a\ll s$.  To be specific, we choose
$a=0.01s$.  This profile shares most of the desirable
properties of (\ref{denprof}); over the wide region where
$a\ll r\ll s$, the density scales like $1/r^2$, and the
scattering angle $\psi$ can still be computed analytically.

The results presented in the table below are for a flat
universe with $H_0=70$ km s$^{-1}$ Mpc$^{-1}$,
$\Omega_m=0.27$ and $\Omega_\Lambda=0.73$.  We present
illustrative parameters for galaxies with $M=10^{11}M_\odot$,
$\sigma=180$ km s$^{-1}$ and $\Omega_g=0.135$, and for
clusters with $M=10^{15}M_\odot$, $\sigma=1000$ km s$^{-1}$
and $\Omega_g=0.0135$, in both cases for sources at
$z_s=1$ and $z_s=2$.  The table lists $D_h$ and $\bar D$, as
well as the difference between $\bar D_{\mathrm{area}}$ and
$\bar D$.

\begin{table}[h]
\begin{footnotesize}
\caption{The angular size and area distances ($\bar D$ and
$D_{{\rm area}}$) to a source at redshift $z_s$ in a
$\Omega_m =$ 0.27, $\Omega_{\Lambda} =$ 0.73 Universe where
matter amounting to a normalized density of $\Omega_g$ exists in
isothermal spheres, each of total mass $M$ and velocity dispersion $\sigma$,
equations (71) through (73), and the rest of the matter is distributed
homogeneously with a normalized density of $\Omega_m - \Omega_g$.
The quantity $\bar D$ represents the average reciprocal magnification
over random directions in the sky, or the average magnification of
randomly selected  sources when the observed images for each source
are unresolved.    The parameter $D_h$, which is the angular size
distance under the scenario of the light path between a small emitter and the
observer passing through only the homogeneous component (i.e. it misses
all the clumps), is also tabulated.}
\begin{center}
\begin{tabular}{ccccccc}
\hline$M$&$\sigma$&$\Omega_g$&$z_s$&$D_h$&$\bar
D$&$D_{\mathrm{area}}-\bar D$\\
($M_\odot$)&(km/s)&&&(Gpc)&(Gpc)&(Mpc)\\ \hline
$10^{11}$& 180&0.135&1&1.74&1.69&13.1\\
$10^{11}$& 180&0.135&2&1.96&1.79&62.6\\
$10^{15}$&1000&0.0135&1&1.74&1.69&0.015\\
$10^{15}$&1000&0.0135&2&1.96&1.79&0.17\\ \hline
\end{tabular}
\end{center}
\end{footnotesize}
\end{table}

The fact that the differences are much larger for galaxies
than for clusters is due to the larger fraction of mass in
the clumps, $\Omega_g$, and to the dependence on $M$ and
$\sigma$.  The relevant parameter is $M/s^2$ or
$\sigma^4/M$, which is larger for galaxies by an order of
magnitude.  At $z_s=1$ the difference between areal and
angular-size distances in not very significant, but by
$z_s=2$, the effect of galaxies is to make the random-source
average 3\% greater than the random-direction one.

\section{Appropriate measures of magnification}

As we noted in the introduction, the appropriate measure of
magnification depends on the type of source being observed,
the mode of observation, and the manner in which the data
are to be interpreted.

Consider first a number count of discrete sources, say
Type-1A supernovae.  Clearly the expected number is
proportional to the area of the source sphere.  However, the
number count can have two meanings.  If the
strong lenses are large --- say galaxy clusters --- so that
the various images of a source are distinct and well
resolved, then we may be interested in the total number of
images, whether or not they come from the same source.  In
that case, we need the \emph{area} distance, $\bar
D_{\mathrm{area}}$; the number of images is proportional to
the \emph{total} area of the source covered as we scan an
area of sky.  On the other hand, if the lenses are smaller
and the various images are not resolved, we would be more
interested in the total number of \emph{distinct} sources. 
In that case, the appropriate measure is the
\emph{angular-size} distance.

A different question we may ask about discrete sources is
their average magnification.  Here we should use the
\emph{random-source} average.  Again, there are two possible
questions we might ask:  What is the average magnification
of each separate image?  Or (if the images are not resolved)
what is the average total magnification of the combined
images?  According to (\ref{mutotav}), the answer to the
second question is exactly the same as in a homogeneous FRW
universe of equaof equal mean density.  On the other hand, the
answer to the first question is obviously less, reduced by
the factor (\ref{muresav}).

The situation with regard to a continuous source such as
the cosmic microwave background (CMB) is different.  If
an intervening lens magnifies a discrete source behind it,
the source appears brighter, because it fills
a larger area of the sky.  This is not the case with the
CMB.  The effect of magnification is that the radiation
we see coming from one patch of sky originates from a
smaller area of the source --- in this case the last
scattering surface.  The surface brightness is
unchanged.  Lensing cannot change the temperature of a
patch of sky, nor the temperature difference between two
different spots.  What it \emph{can} change is the observed
angular separation of these two spots.  In other words, if
there is hot spot on the last scattering surface, it may
appear to us larger or smaller than it would in a
homogeneous universe.

For the CMB, the correlations between the temperatures at
pairs of points are controlled by the distance between
source points rather than the area of source surface
enclosed, so in considering for example the effect on the
position of the acoustic peaks in the angular power spectrum
we should use the \emph{angular-size} distance.  
This is the measure for which the average is precisely the same as in a 
homogeneous universe of equal mean density.  Our conclusion
would not necessarily hold for
\emph{very} small scales, at or below the typical angular separation
of lensed images, though such scales are below the limit of
resolution of current CMB observations.

This does not mean, however, that there could be no observable effect.  
The average magnification is unchanged, but the distribution around the average 
could be markedly affected with a separation between the mode
and the mean, \citet{LMa05}.  Moreover,
the acoustic peaks may be observably broadened, \citet{LMb05}.
One source of skewness in the distribution operates
on an angular scale such that the
average region contains less than one clump.  In that case,
when we choose a random area of sky of the prescribed size,
it is very likely to contain no clumps at all, in which case
it will be demagnified.  On the other hand in those rare
cases when it does contain a clump, there will be a
comparatively large magnification.  So the peak may be
shifted to a smaller scale.  There has already been
considerable discussion of this issue in the literature; see
for example \citep{ZalS99,BarS01}. 

\section{Discussion}

We have seen that so long as the weak-lensing condition is
well satisfied, the average magnification or reciprocal
magnification (in whatever sense) is essentially
the same as in a homogeneous universe of equal mean (or pre-clumping)
density, \emph{provided} of course that our assumptions are
correct, namely that the clumps are uncorrelated and well
separated and that the gravitational effect of each can be
treated in a Newtonian approximation.  When the clumps are
\emph{not} well separated, then it may no longer be
legitimate to assume that their contributions simply add,
though we see no reason why that should not be true.  We
have also ignored any time-dependence in the Newtonian
potential, so the clumps must not be moving too rapidly, and
must be small enough for the integrated Sachs--Wolfe effect
to be negligible.  Probably the most significant shortcoming
of our approach is the assumption that the clumps are
uncorrelated.  Although we have considered explicitly only
one size, there would be no real difficulty in including
clumps of many sizes.  But the effect of correlation
between clumps of different sizes might well be important.

When we approach the strong-lensing regime, on the other
hand, the various averages are not the same, and in
particular we must distinguish angular-size and areal
distances.  The random-direction angular-size average of the
reciprocal magnification $\langle 1/\mu\rangle_{\mathrm{rd}}$
is just the same as in a homogeneous universe of equal
mean density, and as a corollary, the same is true of the
random-source average of the \emph{total} magnification of
unresolved images, $\langle
\mu^{\mathrm{tot}}\rangle_{\mathrm{rs}}$.  However, the
random-source average of the magnification of \emph{resolved}
images, $\langle
\mu^{\mathrm{resolved}}\rangle_{\mathrm{rs}}$ may be
significantly different, especially for the most distant
sources.  The difference between $\bar D_{\mathrm{area}}$
and $\bar D$ increases rapidly with increasing $z$.

\section*{Appendix. Average of $A$}

In this appendix, we present a proof of the relation
(\ref{detAbarD}) between $\langle A\rangle$ and $\bar
D$.

Let us consider a cylinder of fixed radius $R$ centered on
our fiducial geo\-desic, and extending out to $\lambda_s$. 
We choose $R$ sufficiently large that clumps outside the
cylinder have negligible effect; the value of $R$ will drop
out of the final answer.  It is useful to consider in more
detail the process of averaging over clumps.  Clearly by
(\ref{no}), the probability that \emph{no} clumps are found
within the cylinder is
 \begin{equation}
P_0(\lambda_s)=\exp\left[-n_0\pi
R^2\int_0^{\lambda_s}(1+z)^4d\lambda\right].
 \end{equation}
The probability that exactly $N$ clumps are found in the
cylinder, at positions $\{(\lambda_j,-\vec x_j),
j=1,2,\dots N\}$ is 
 \begin{equation}
P_0(\lambda_s)n_0^N\prod_{j=1}^N
(1+z_j)^4d\lambda_j d^2\vec x_j.
\label{meas}
 \end{equation}

Now what is the value of ${\cal A}(\lambda_s)$ if $N$ clumps
are present at these positions?  Of course, if $N=0$, then
${\cal A}(\lambda_s)=D_h(0,\lambda_s){\bf 1}$.  If there is
just one clump, at $(\lambda_1,-\vec x_1)$, its deviation
effect is added, so we get
 \begin{equation}
{\cal A}(\lambda_s)=D_h(0,\lambda_s){\bf 1}+
D_h(0,\lambda_1){\cal K}(\vec x_1)
D_h(\lambda_1,\lambda_s).
 \end{equation}
For $N=2$, there are four terms, these two plus two more
in which ${\cal K}(\vec x_2)$ also appears.  In general, we
can write
 \begin{equation}
{\cal A}(\lambda_s)=\sum_I{\cal A}_I,
\label{sumAI}
 \end{equation}
where the sum is over all subsets $I$ of $\{1,2,\dots
N\}$.  Here, if $I=\{i_1,i_2,\dots i_P\}$, then
 \begin{equation}
{\cal A}_I(\lambda_s)=\prod_{j=0}^P
D_h(\lambda_{i_j},\lambda_{i_{j+1}})
\prod_{j=1}^P{\cal K}(\vec x_{i_j}),
\label{AI}
 \end{equation}
where $\lambda_{i_0}=0$ and $\lambda_{i_{P+1}}=\lambda_s$.

We can now recover our expression for $\langle{\cal
A}(\lambda_s)\rangle$.  To do this we integrate
(\ref{sumAI}) over the measure (\ref{meas}) and then sum
over $N$.  Now when we integrate over $\vec x_i$, then if
$i\in I$, by (\ref{intK}) we recover a factor of $-4\pi
GM{\bf 1}$, whereas if $i\notin I$, we get simply $\pi
R^2$.  It is then convenient to sum separately over the
number of elements $P$ in the selected subset $I$, and the
number in the complement, $Q=N-P$.  It is easily seen
that the sum over $Q$ gives an exponential that precisely
cancels the factor $P_0(\lambda_s)$.  Thus, again using
(\ref{no}), we obtain
 \begin{eqnarray}
\langle{\cal A}(\lambda_s)\rangle&=&{\bf 1}
\sum_{P=0}^\infty\left(-\frac{3}{2}H_0^2\Omega_g\right)^P
\int_0^{\lambda_s}d\lambda_1
\int_{\lambda_1}^{\lambda_s}d\lambda_2\dots
\int_{\lambda_{P-1}}^{\lambda_s}d\lambda_P\nonumber\\
&&\prod_{j=1}^P(1+z_j)^4\prod_{j=0}^P
D_h(\lambda_j,\lambda_{j+1}),
\label{aveA}
 \end{eqnarray}
again with $\lambda_0=0$ and $\lambda_{P+1}=\lambda_s$.  It
is easy to see that this is precisely the perturbation
solution in powers of $\Omega_g$ of (\ref{Addot}), so this
again proves that $\langle{\cal A}(\lambda_s)\rangle=\bar
D(\lambda_s){\bf 1}$.  (Note that the required extra factor
of $(1+z_j)$ comes from the initial condition (\ref{Dh2ic})
for $D_h(\lambda_j,\lambda_{j+1})$.)

Next we turn to the computation of
$\langle A\rangle=\langle\det{\cal A}\rangle$.  Suppose
as before that there are $N$ clumps within our cylinder. 
Then from (\ref{sumAI}), using the identity 
$\det{\cal A}=\frac{1}{2}[({\rm tr}{\cal A})^2-{\rm
tr}({\cal A}^2)]$ for $2\times2$ matrices, we have
 \begin{equation}
\det{\cal A}(\lambda_s)=\frac{1}{2}\sum_I\sum_J
[({\rm tr}{\cal A}_I)({\rm tr}{\cal A}_J)
-{\rm tr}({\cal A}_I{\cal A}_J)],
\label{detAIJ}
 \end{equation}
where $I$ and $J$ are arbitrary subsets of
$\{1,2,\dots,N\}$, and ${\cal A}_I$, ${\cal A}_J$ are
given by (\ref{AI}).

Now consider the effect of the integrations over $\vec
x_i$.  As before, if $i\notin I$ and $i\notin J$, the
integral gives simply $\pi R^2$, and the sum over all such
contributions generates an exponential that cancels
$P_0(\lambda_s)$.  Next, consider the indices that belong
to one of the two subsets only, say $i\in I$ but $i\notin
J$.  Then ${\cal K}(\vec x_i)$ appears only once in each
term of (\ref{detAIJ}), so we can immediately perform the
integral, and replace ${\cal K}(\vec x_i)$ with a factor
$-4\pi GM{\bf 1}$.  Lastly, we are left with the indices,
say $\{i_1,i_2,\dots i_S\}$ that are common to both $I$ and
$J$.  They contribute a factor
 \begin{eqnarray}
\frac{1}{2}\int d^2\vec x_{i_1}\dots d^2\vec x_{i_S}&&
\!\!\!\!\!\!\!\!
\left(\{{\rm tr}[{\cal K}(\vec x_{i_1})\dots{\cal K}(\vec
x_{i_S})]\}^2-{\rm tr}\{[{\cal K}(\vec x_{i_1})\dots{\cal
K}(\vec x_{i_S})]^2\}\right)\nonumber\\
&&=\int d^2\vec x_{i_1}\dots d^2\vec x_{i_S}\det
[{\cal K}(\vec x_{i_1})\dots{\cal K}(\vec x_{i_S})].
 \end{eqnarray}
But now in virtue of (\ref{intdetK}), this expression
vanishes unless $S=0$.

So finally we are left with a sum only over disjoint sets
$I$ and $J$.  When we have performed the integrations over
$\vec x_i$, all the matrices are proportional to $\bf 1$,
so the trace merely gives a factor of $2$ which cancels the
$\frac{1}{2}$ in (\ref{detAIJ}).  The remaining factors,
for each of the two sums separately, are identical with
those in (\ref{aveA}), so we find
 \begin{equation}
\langle A(\lambda_s)\rangle=[\bar D(\lambda_s)]^2.
 \end{equation}
This concludes the proof.

\section*{Acknowledgements}

We are indebted to R.D.~Blandford and P.J.E.~Peebles for
helpful discussions.

\end{document}